\documentclass[%
 reprint,
superscriptaddress,
 amsmath,amssymb,
 aps,
prl,
]{revtex4-1}

\usepackage{graphicx}
\usepackage{epstopdf}
\usepackage{textcomp}
\usepackage{dcolumn}
\usepackage{bm}
\usepackage{gensymb}
\usepackage{upgreek}
\newcommand{\BaNi}{BaNi$_2$As$_2$}
\newcommand{\BaNiCo}{Ba(Ni$_{1-x}$Co$_{x}$)$_2$As$_2$}
\newcommand{\Tc}{$T_c$}
\newcommand{\Hc}{$H_{c2}$}

\newcommand{\etal}{\textit{et al. }}

\begin{document}

\preprint{APS/123-QED}

\title{Evolution of Structure and Superconductivity in \BaNiCo}

\author{Chris Eckberg}
\email{eckbergc@umd.edu}
\affiliation{Center for Nanophysics and Advanced Materials, Department of Physics, University of Maryland, College Park, Maryland 20742, USA}

\author{Limin Wang}
\affiliation{Center for Nanophysics and Advanced Materials, Department of Physics, University of Maryland, College Park, Maryland 20742, USA}
\author{Halyna Hodovanets}
\affiliation{Center for Nanophysics and Advanced Materials, Department of Physics, University of Maryland, College Park, Maryland 20742, USA}
\author{Hyunsoo Kim}
\affiliation{Center for Nanophysics and Advanced Materials, Department of Physics, University of Maryland, College Park, Maryland 20742, USA}
\author{Daniel J. Campbell}
\affiliation{Center for Nanophysics and Advanced Materials, Department of Physics, University of Maryland, College Park, Maryland 20742, USA}
\author{Peter Zavalij}
\affiliation{Department of Chemistry, University of Maryland, College Park, Maryland 20742, USA}
\author{Phil M. Piccoli}
\affiliation{Department of Geology, University of Maryland, College Park, Maryland 20742, USA}

\author{Johnpierre Paglione}
\email{paglione@umd.edu}
\affiliation{Center for Nanophysics and Advanced Materials, Department of Physics, University of Maryland, College Park, Maryland 20742, USA}

\date{\today}

\begin{abstract}

The effects of Co-substitution on \BaNiCo~(0 $\leq x\leq$ 0.251) single crystals grown out of Pb flux are investigated via transport, magnetic, and thermodynamic measurements. \BaNi~exhibits a first order tetragonal to triclinic structural phase transition at $T_s$ = 137 K upon cooling, and enters a superconducting phase below \Tc~= 0.7 K. The structural phase transition is sensitive to cobalt content and is suppressed completely by $x$ $\geq$ 0.133. The superconducting critical temperature, \Tc, increases continuously with $x$, reaching a  maximum of \Tc~= 2.3 K at the structural critical point $x$ = 0.083 and then decreases monotonically until superconductivity is no longer observable well into the tetragonal phase. In contrast to similar \BaNi~substitutional studies, which show an abrupt change in \Tc~at the triclinic-tetragonal boundary that extends far into the tetragonal phase, \BaNiCo~exhibits a dome-like phase diagram centered around the first-order critical point. Together with an anomalously large heat capacity jump $\Delta$\textit{C$_e$}/$\gamma$T $\sim$ 2.2 at optimal doping, the smooth evolution of \Tc~ in the \BaNiCo~ system suggests a mechanism for pairing enhancement other than phonon softening.

\end{abstract}

\maketitle

\section{Introduction}

High temperature superconductivity in Fe-based compounds has taken on immense research interest since their discovery in 2008 \cite{KamiharaFeSCs, PaglioneRev,JohnstonRev, LumsdenRev}. Of these compounds, BaFe$_2$As$_2$ has been among the most extensively studied, largely due to availability of sizable high quality single crystals. BaFe$_2$As$_2$ is an antiferromagnet (AFM) with $T_N$ = 135 K \cite{Rotter}. AFM order is closely linked to both electronic nematicity \cite{Fernandes} and structural symmetry breaking from tetragonal to orthorhombic\cite{Rotter}. Magnetic and structural transitions present in pure BaFe$_2$As$_2$ are sensitive to chemical substitution and physical pressure, and a dome-like superconducting phase emerges with their suppression \cite{Canfield}.  While substitution on all three ionic sites has been observed to stabilize high \Tc~superconductivity, the choice of substituant site strongly influences the ensuing superconducting phase. For instance, electron doping with Ni and Co substitution for Fe induces fully gapped superconductivity while isoelectronic substitution of P on the As site produces a nodal superconducting phase \cite{Stewart,Co-nodeless, Ni-nodeless, P-nodal}. Superconductivity in all of these series however is believed to be closely linked to phase criticality; specifically, the competition and cooperation between nematic and magnetic phases and superconducting pairing.  

\BaNi~crystallizes in the same tetragonal ThCr$_2$Si$_2$ structure (space group \textit{I4/mmm}) as BaFe$_2$As$_2$ and similarly undergoes a structural distortion at approximately 135 K \cite{Ronning}. However, in \BaNi~the structural distortion is between a high temperature tetragonal and low temperature triclinic, rather than orthorhombic, symmetry, and has no associated magnetic order \cite{Sefat, Neutron}. Rather, theoretical work has suggested that the zig-zag chain structure in the triclinic distortion is driven by orbital ordering, explaining the lack of magnetic order \cite{Zigzag}.

\BaNi~also displays bulk superconductivity below \Tc~= 0.7 K \cite{Ronning}, suggested to be conventional BCS-type in nature with a fully-gapped $s$-wave order parameter symmetry \cite{Kurita, Subedi}. Superconductivity in \BaNi~is widely thought to be distinct from the unconventional sign-changing, $s^\pm$, order parameter of the iron-based high-\Tc~ superconductors \cite{s+-,PaglioneRev}, such as in Ba(Fe$_{1-x}$Ni$_{x}$)$_2$As$_2$ 0.02 $\leq x\leq$ 0.08 \cite{Fe-NiPD}. Electronic structure calculations suggest \BaNi~should not host an $s^\pm$ state, as any nodal planes would necessarily intersect the Fermi Surface due to its complexity \cite{NiRev}, and the heat capacity and thermal conductivity data of \BaNi~has been well fit to a BCS $s$-wave model \cite{Kurita}. Despite the distinctions from its iron-based counterpart, previous substitutional studies in Ba(Ni$_{1-x}$Cu$_{x}$)$_2$As$_2$ \cite{KudoCu} and BaNi$_2$(As$_{1-x}$P$_{x}$)$_2$ \cite{KudoP} have found an abrupt, strong enhancement of \Tc~from 0.7 K to 3.3 K upon suppression of the triclinic phase \cite{KudoCu, KudoP}, with strengthened pairing attributed to a soft phonon mode at the first-order structural phase boundary. The enhanced \Tc~value in the tetragonal phase of BaNi$_2$(As$_{1-x}$P$_{x}$)$_2$ extends to the $x$ = 1 end-member BaNi$_2$P$_2$ \cite{BaNiP}, suggesting the enhancement is rooted in the tetragonal structure itself.

The recent discovery of charge density wave (CDW) emerging near the structural transition in \BaNi~\cite{Abbamonte} raises new questions about pairing in this system, in particular the possibility of a more complicated relationship between superconductivity and structural criticality in \BaNi. Here we report the physical properties of Co-substituted \BaNi~single crystals, showing that the low temperature triclinic phase is smoothly suppressed with cobalt substitution concomittant with a continuous enhancement of \Tc~upon approach to the zero-temperature structural phase boundary. We find that, in contrast to other reported  \BaNi~substitutional studies, and in a manner reminiscent of similar work in BaFe$_2$As$_2$, \BaNiCo~exhibits a strong enhancement of \Tc~in both the triclinic and tetragonal low temperature phases, with suppression away from the structural critical point suggesting a Cooper pairing enhancement reminiscent of superconductivity emerging from quantum criticality.

\begin{figure}
    \centering
    \includegraphics[width=0.47\textwidth]{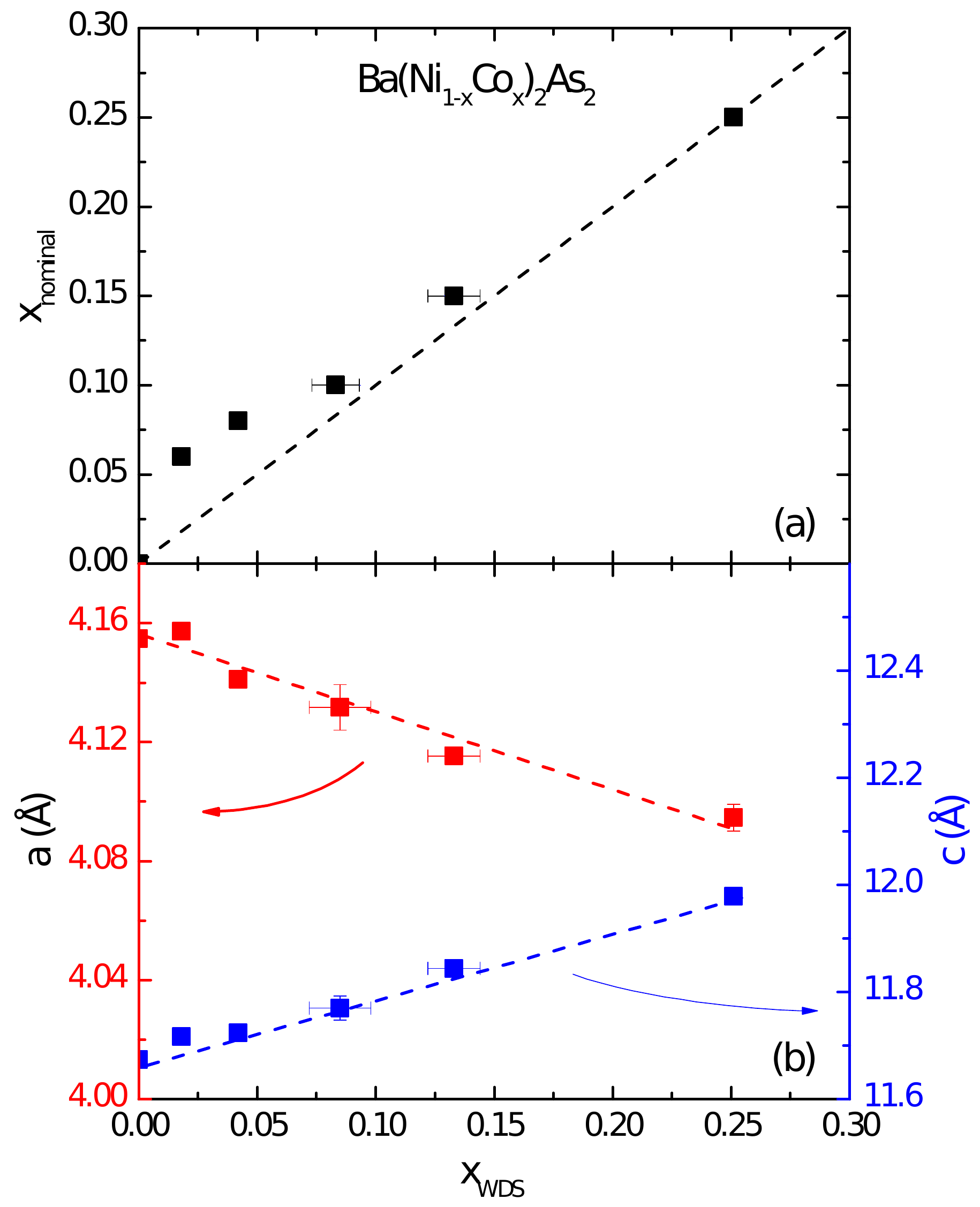}
    \caption{Structural and chemical characterization of \BaNiCo~single crystals. (a) WDS chemical composition characterization for \BaNiCo~single crystals. $x_{wds}$ = $x_{nominal}$ curve represented by black dashed line. (b) Lattice parameters in \BaNiCo~series collected at 250 K. Data show a strongly linear evolution in both $a$ and $c$ axis length through $x$ = 0.251.}
    \label{fig:Figure1}
\end{figure}

\begin{figure}
    \centering
    \includegraphics[width=0.47\textwidth]{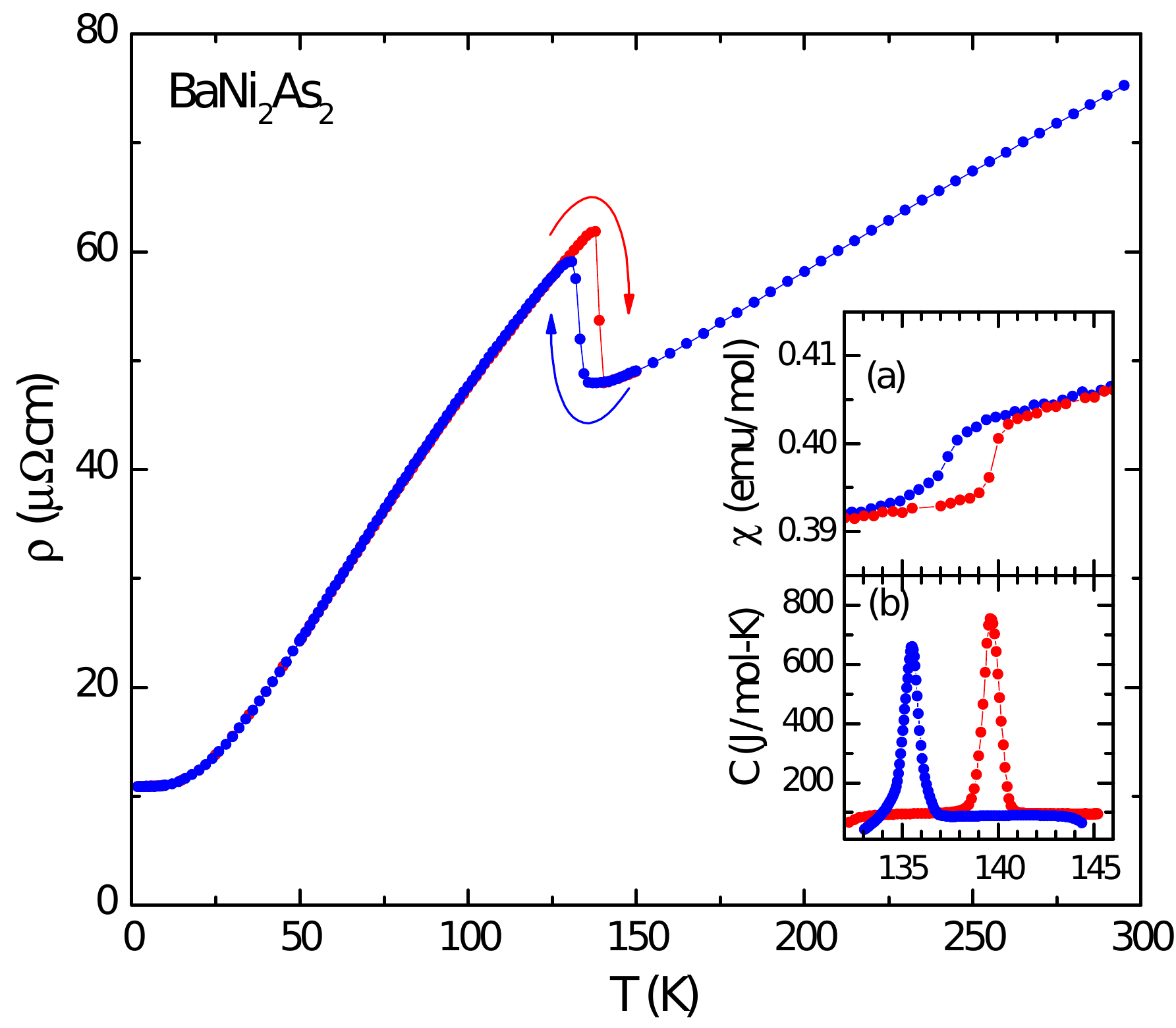}
    \caption{Characterization of the structural transition in \BaNi. Resistivity of pure \BaNi~single crystals shown in main figure. Inset a (b) displays hysteretic magnetization (heat capacity) when warming and cooling through the structural transition.}
    \label{fig:Figure2}
\end{figure}

\section{Experimental Methods}

\begin{table}
\caption{\label{tabl1} \BaNiCo~crystallographic data determined through single-crystal x-ray diffraction. All data were collected at 250 K.}
\footnotesize\rm
\begin{ruledtabular}
\begin{tabular}{llll}
$x$&0&0.083&0.133\\
\hline
Crystal system&Tetragonal&Tetragonal&Tetragonal\\
Space group&I4/mmm&I4/mmm&I4/mmm\\
$a$($\mathrm{\AA}$)&4.144(2)&4.1256(5)&4.1140(7)\\
$b$($\mathrm{\AA}$)&4.144(2)&4.1256(5)&4.1140(7)\\
$c$($\mathrm{\AA}$)&11.656(6)&11.7486(15)&11.827(2)\\
$V^3$($\mathrm{\AA}^3)$&200.2(2)&199.97(5)&200.17(8)\\
Reflections&1737&1705&1776\\
R$_1$&0.0140&0.0179&0.0156\\
Atomic parameters:&&&\\
Ba&2$a$ (0,0,0)&2$a$ (0,0,0)&2$a$ (0,0,0)\\
Ni/Co&4$d$ (0,1/2,1/4)&4$d$ (0,1/2,1/4)&4$d$ (0,1/2,1/4)\\
As&4$e$ (0,0,$z$)&4$e$ (0,0,$z$)&4$e$ (0,0,$z$)\\
$z$&0.34726(6)&0.34785(7)&0.34812(6)\\
Bond lengths ($\mathrm{\AA}$):&&\\
Ba-As($\mathrm{\AA}$)&3.4288(15)&3.4213(6)&3.4189(6)\\
Ni/Co-As($\mathrm{\AA}$)&2.3619(11)&2.3615(5)&2.3618(5)\\
As-As($\mathrm{\AA}$)&3.560(71)&3.575(0)&3.592(5)\\
Bond angles (deg):&&&\\
As-Ni/Co-As&103.32(2)&103.710(16)&103.971(15)\\
As-Ni/Co-As&122.63(5)&121.74(4)&121.14(3)\\
\end{tabular}
\end{ruledtabular}
\end{table}

\BaNiCo~crystals were grown out of Pb-flux using a solution growth technique originally reported by Ronning, \etal\cite{Ronning}. Crystals formed as shiny, thin platelets, with typical dimensions of 0.5 mm $\times$ 0.5 mm $\times$ 0.05 mm with a high observed residual resistivity ratio RRR = 10 that exceeded previous reports, as well as our own self-flux grown samples. The typically small crystal sizes were prohibitive for thermodynamic and magnetic measurements. To circumvent this issue, larger \BaNi~ crystals with dimensions of 2 mm $\times$ 2 mm $\times$ 0.5 mm were also synthesized out of NiAs self-flux \cite{Sefat} and were used for characterization of the structural transition.

Elemental composition in substituted samples was determined using wavelength dispersive spectroscopy (WDS). Crystal properties within a growth show minimal variation, while WDS gives variability in \BaNiCo ~Co concentrations of $\Delta x \leq$ 0.01 for crystals pulled from the same growth. Variation between the nominal \textit{x} concentration versus the one obtained from WDS is shown in Fig. 1(a).

Structural data were collected on single crystals in a Bruker APEX-II CCD system equipped with a graphite monochromator and a MoK$_\alpha$~sealed tube ($\lambda$ = 0.71073 \AA), and were refined using the Bruker SHELXTL Software Package. Crystallographic information collected in the tetragonal phase (250 K) are included in Table 1 for several representative $x$ values. Atomic positions evolve monotonically across the phase diagram, while low conventional residual values (R$_1$) confirm high crystal quality. A continuous decrease in $a$ and increase in $c$ lattice parameters with increasing Co concentration is observed across all measured samples, as shown in Fig. 1(b). 

Standard density functional theory calculations for pure \BaNi~were conducted using the WIEN2K \cite{Wein2k} implementation of the full potential linearized augmented plane wave method in the local density approximation. The k-point mesh was taken to be 11 $\times$ 11 $\times$ 11, with lattice constants taken from our experimental measurements. Supercell calculations were implemented for Co-substituted cases (i.e., Ba$_4$Ni$_7$CoAs$_8$ for $x$ = 0.125 and Ba$_2$Ni$_3$CoAs$_4$ for $x$ = 0.250), and resultant electronic structures unfolded via recently developed first-principles unfolding methods \cite{Unfolding}.

Transport, heat capacity, and $ac$ magnetic susceptibility data were taken using Quantum Design Physical Property Measurement System\textsuperscript{\textregistered} (PPMS-14T) and DynaCool\textsuperscript{TM} (DC-14T) systems. An environment between 1.8 and 300 K was used in each system. Heat capacity and transport measurements were extended down to 400 mK using a Quantum Design Helium-3 refrigerator option compatible with the PPMS. In-plane transport data were taken using a four wire configuration. Au wires were attached to cleaved, or polished when necessary (to remove Pb contamination) single crystals using DuPont 4929N silver paste. Single crystal \textit{ac} magnetic susceptibility was measured using a homemade coil \cite{Coil}. \textit{Ac} magnetic susceptibility measurements between 0.1 and 3 K were taken with the coil mounted on a Quantum Design Adiabatic Demagnetization Refrigerator insert for the PPMS. Data were taken at a frequency of 19.997 kHz, in an ac field with approximate amplitude of 0.25 Oe.

Heat capacity measurements were taken with a relaxation technique fit to a dual time constant model. The background heat capacity of the platform and grease was measured first and subtracted from the final result. Experiments on Co substituted samples were complicated by small crystal sizes ($<$ 0.1 mg). To circumvent this issue, heat capacity measurements were taken on collections of several samples pulled from the same growth. Sharp anomalies at the structural transitions in measurements taken on these collections of crystals, along with the high degree of growth homogeneity determined through WDS, suggest minimal error in heat capacity data due to collection averaging. The heat capacity data across the first-order structural transition of BaNi$_2$As$_2$ was measured by establishing $\Delta T$ of 15 K at 130 K\cite{QD}. Data were collected over 4$\tau$ measuring time (about 2.5 min). Single slope method \cite{QDHC} was used to calculate the heat capacity that is shown in the inset (a) to Fig. 2.

DC magnetic susceptibility measurements were carried out in a Quantum Design Magnetic Property Measurement System (MPMS) SQUID magnetometer.

\begin{figure}
    \centering
    \includegraphics[width=0.47\textwidth]{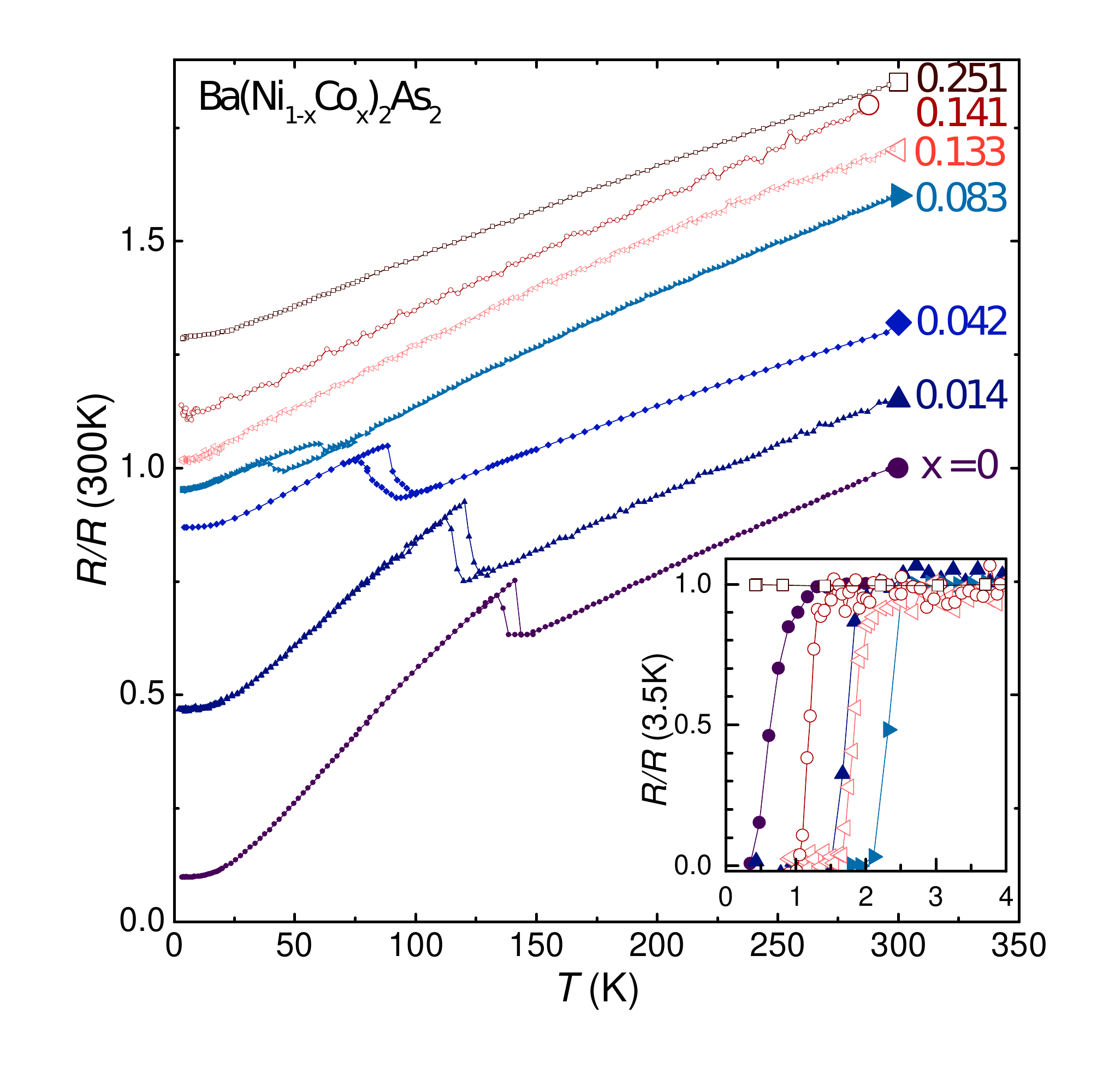}
    \caption{Transport measurements in \BaNiCo~series. Main figure displays resistance normalized to room temperature value, and vertically offset for clarity. Data show clear suppression of anomalies associated with the structural transition, which vanishes by $x$ = 0.133. Inset displays low temperature resistance normalized to 3.5 K value. Samples display clear enhancement of \Tc~when approaching structural phase boundary. Data plotted in blue ($x$ $\leq$ 0.083) feature a low temperature resistance anomaly consistent with the triclinic structural distortion. Curves plotted in red remain tetragonal down to the lowest measured temperature.}
    \label{fig:Figure3}
\end{figure}

\begin{figure}
    \centering
    \includegraphics[width=0.47\textwidth]{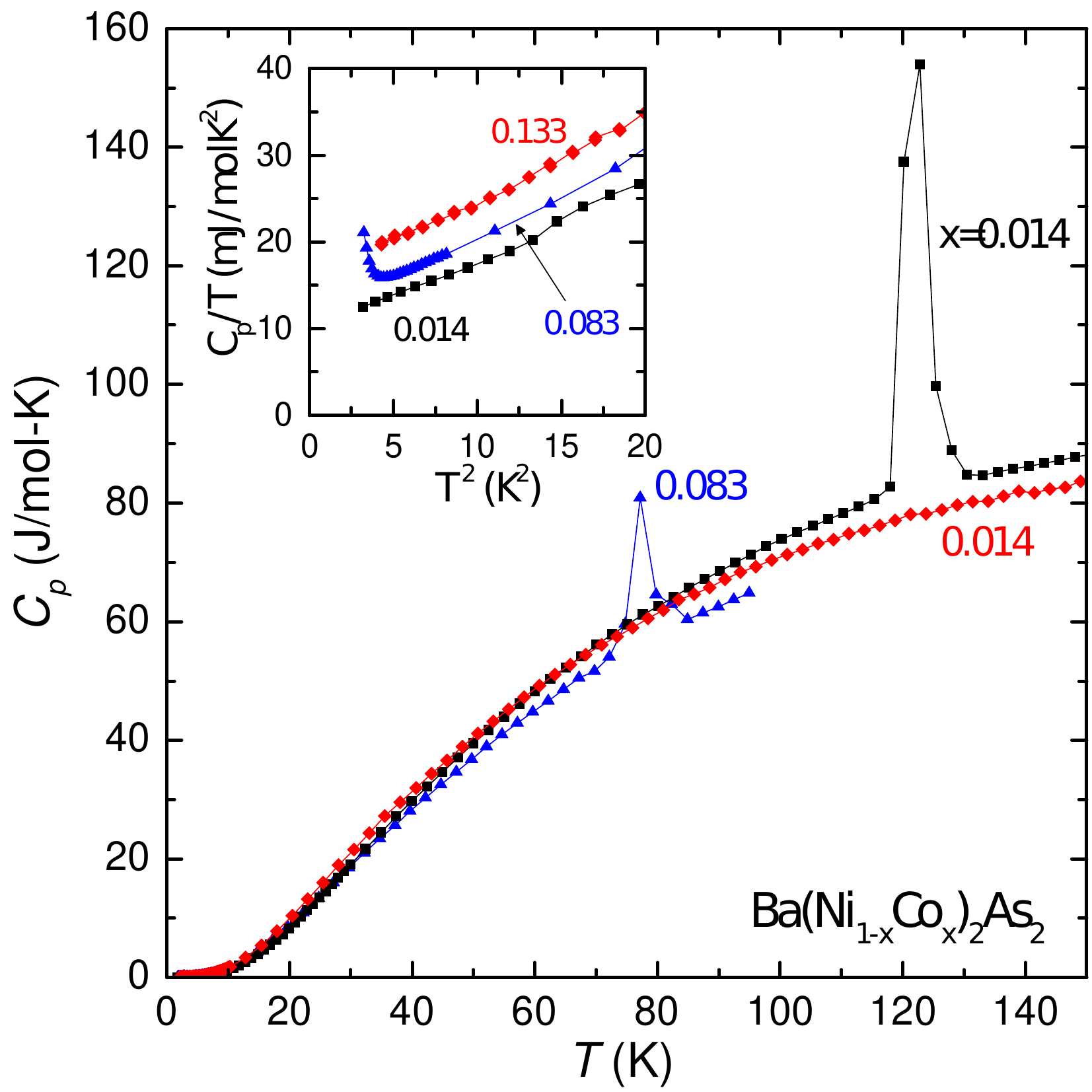}
    \caption{Heat capacity measurements collected on warming in \BaNiCo~crystals. Anomalies in main figure indicative of structural transition. Inset displays low temperature $C_p$/$T$ data plotted versus temperature squared.}
    \label{fig:Figure4}
\end{figure}

\begin{figure}
    \centering
    \includegraphics[width=0.47\textwidth]{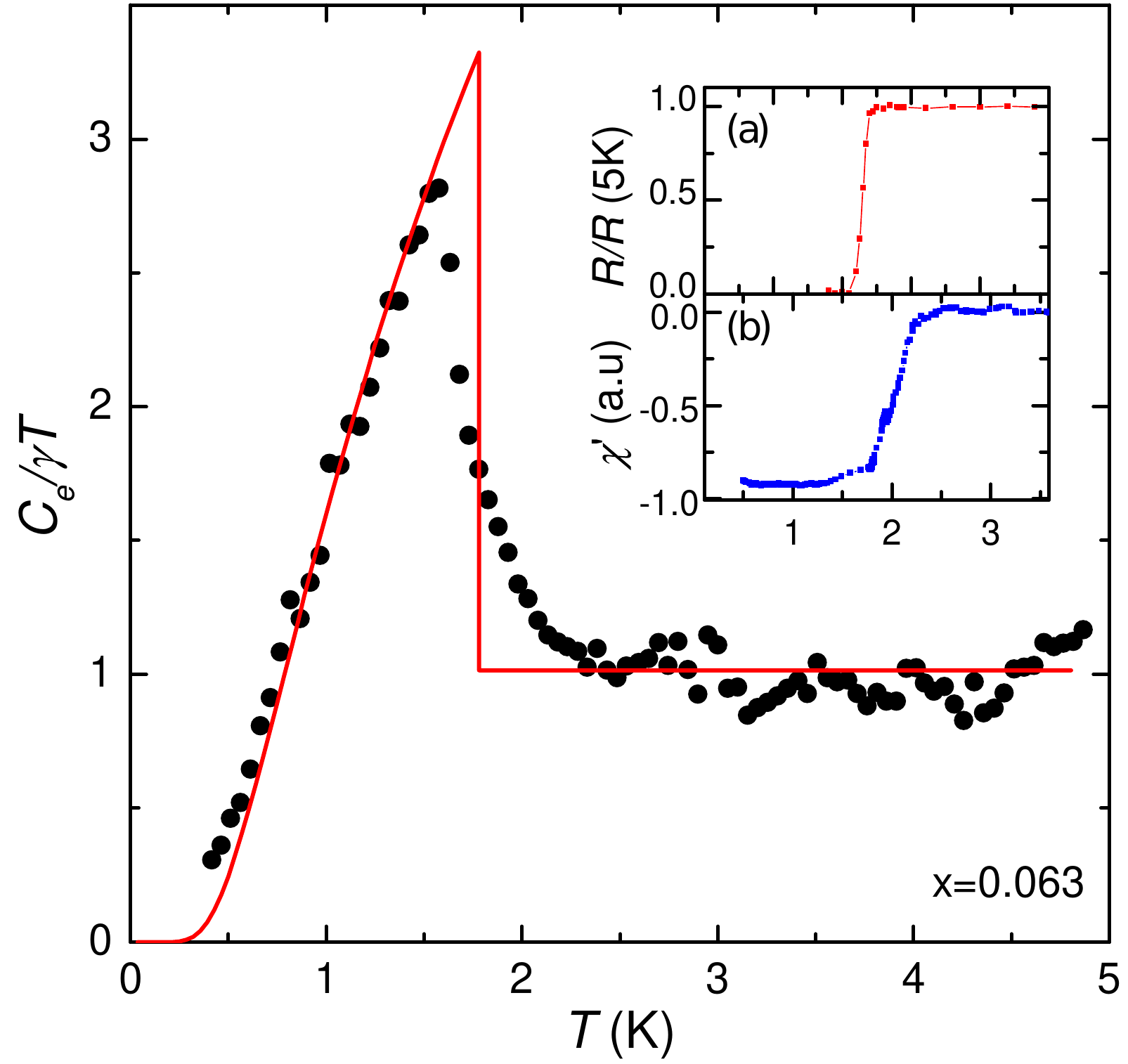}
    \caption{Measurements of superconducting transition in sample with near optimal substitution $x$ = 0.063. Electronic heat capacity ($C_e$) was determined by subtracting the phonon contribution ($\beta$T$^3$) from the total heat capacity (main figure). Red curve is the $\alpha$-model predictions for a BCS superconductor \cite{Johnston} ($\alpha$ = 1.764) scaled by a constant multiple to match the data. Inset displays superconducting transition measured via four terminal resistance (a) and the real part of $ac$ susceptibility measured using a homemade coil (b).}
    \label{fig:Figure5}
\end{figure}

\begin{figure}
    \centering
    \includegraphics[width=0.47\textwidth]{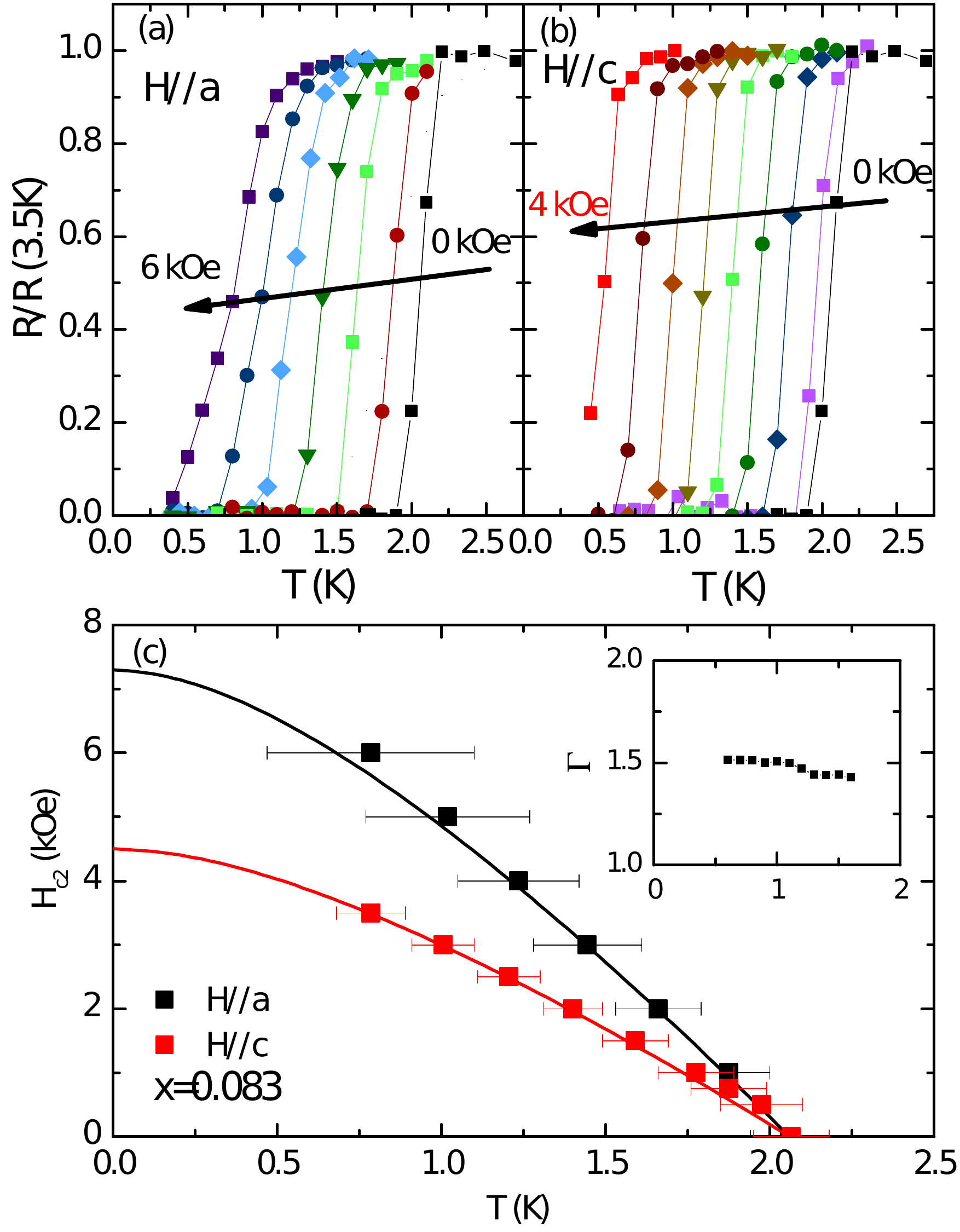}
    \caption{\Hc~data collected on an optimally substituted sample of $x$ = 0.083. Isomagnetic resistance data collected with field parallel to crystal a-axis (a) and parallel to crystal c-axis (b). Data collected parallel to the c-axis were taken in 0.5 kOe increments (0, 0.5, 1, 1.5, 2, 2.5, 3, 3.5 and 4 kOe), while measurements with H $\parallel$ a were taken in 1 kOe increments (0, 1, 2, 3, 4, 5, and 6 kOe). (c) \Hc~vs \Tc~phase diagram in parallel and perpendicular configuration.Data points were taken at the midpoint of the resistive transition, and error bars represent the range wherein resistance is between 90\% and 10\% the normal state value. Curves are generated for a dirty BCS superconductor using the model developed by Werthamer, Helfand, and Hohenberg \cite{WHH}. Inset shows upper critical field anisotropy, $\Gamma$, determined using the midpoint criteria.}
    \label{fig:Figure6}
\end{figure}

\section{Results}
The electrical resistivity of BaNi$_2$As$_2$ is presented in Fig. 2, showing a pronounced hysteresis in the data collected on warming and cooling due to the strongly first-order tetragonal to triclinic structural transition. This hysteresis is also observed in the magnetic susceptibility (see inset (a) to Fig. 2) and heat capacity (see inset (b) to Fig. 2). Figure 3 displays the evolution of the hysteretic region in \BaNiCo~as measured by resistivity on heating and cooling. The hysteresis, and by extension the structural distortion, is observed throughout the range of the triclinic phase and is quickly suppressed with increasing $x$. This observation is consistent with the evolution of the heat capacity anomaly shown in Fig. 4, which also is absent by $x$ = 0.133. The low temperature heat capacity displayed in the inset to Fig. 3 shows no dramatic changes in Debye temperature or Sommerfield coefficient for the reported Co concentrations. The extracted Debye temperatures are $\Theta$$_D$ = 236 K, 218 K, and 225 K for $x$ = 0.014, 0.083, and 0.133 respectively. Pure \BaNi~was observed to have a Debye temperature of 250 K, consistent with previous work \cite{Sefat}.

Despite large changes in low temperature structure, superconductivity surprisingly evolves continuously in the \BaNiCo~series (see Fig. 3 inset), with a fast enhancement in \Tc~upon cobalt substitution, rising from \Tc~= 0.7 K at $x$ = 0 to 1.7 K with just 1.4 \% cobalt substitution for nickel. \Tc~continues to increase with $x$ in the triclinic phase, eventually exhibiting a maximum of 2.3 K at $x$ = 0.083 and then gradual decreasing until entirely absent by $x$ = 0.251. 
Fig.~5 presents heat capacity (main), transport (a) and $ac$ magnetization (b) measurements of the superconducting transition in the same single-crystal sample (crystal dimensions of 0.67 mm $\times$ 0.83 mm $\times$ 0.067 mm) with $x$ = 0.063. Balancing the entropy in the observed heat capacity jump yields a \Tc~of 1.8 K for this sample. The red curve in Fig. 5 is the $\alpha$ model prediction of heat capacity for a single band BCS superconductor. This curve has been scaled by a constant value of 1.35 to match the observed heat capacity jump. This model describes data well near \Tc, and deviations at low temperatures may be due to nuclear Schottky contributions as observed in the pure compound \cite{Kurita}. The modeled heat capacity jump $\Delta$\textit{C$_e$}/$\gamma$T is $\sim$2.2, well above the BCS limit of 1.43, indicating strongly coupled superconductivity at this Co concentration. This value is consistent with previous reports of enhanced normalized heat capacity jumps of approximately 1.9 in both Ba(Ni$_{1-x}$Cu$_{x}$)$_2$As$_2$ and BaNi$_2$(As$_{1-x}$P$_{x}$)$_2$ \cite{KudoCu, KudoP} and greatly exceeds the near-BCS value observed in pure \BaNi~\cite{Kurita}. While previous work on Cu- and P-substituted \BaNi~ suggested that the enhancement in the tetragonal phase was consistent with a phonon softening picture, this is not the case here, as the enhancement occurs in the triclinic phase and the Debye frequency exhibits little change through the entire Co substitution range as noted above.

Both superconductivity and the structural transition in optimally substituted $x$ = 0.083 samples were observed to be of bulk origin, as each manifested itself in anomalies in measured heat capacity (see Fig. 4 main and inset). Figure 6 shows the evolution of upper critical field in these optimally substituted samples, which exhibit an approximately three fold enhancement compared to the pure compound. As reported in \BaNi, superconductivity in optimally substituted \BaNiCo~is more robust when field is applied parallel to the crystal plane. At higher fields, resistance curves taken in this orientation begin to broaden, while data taken with field along the $c$ axis remain sharp over all measurements. \Hc~anisotropy, $\Gamma$, remains virtually constant at all temperatures, with $\Gamma$ = 1.50 slightly below the value 2.1 reported for the pure compound \cite{Ronning}. 

\begin{figure}
    \centering
    \includegraphics[width=0.47\textwidth]{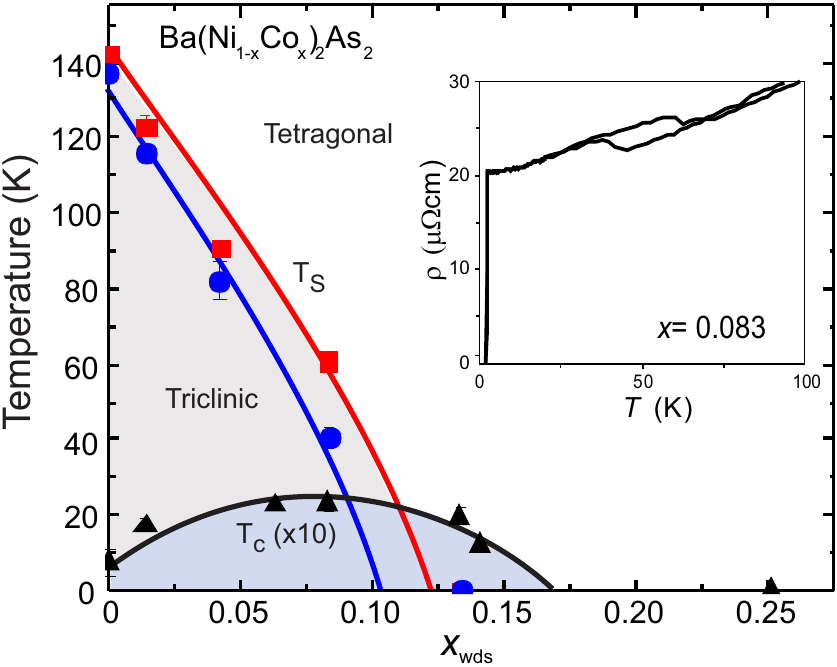}
    \caption{Phase diagram for \BaNiCo~system gathered from transport data. Structural and superconducting critical temperatures were both determined by the midpoint of each resistive transition. Superconducting \Tc~is scaled by a factor of 10 to improve clarity. Inset displays transport data for optimally substituted, $x$ = 0.083, samples featuring both clear enhancement in \Tc~and structural transition anomaly.}
    \label{fig:Figure7}
\end{figure}

\section{Discussion}

\begin{figure}
    \centering
    \includegraphics[width=0.47\textwidth]{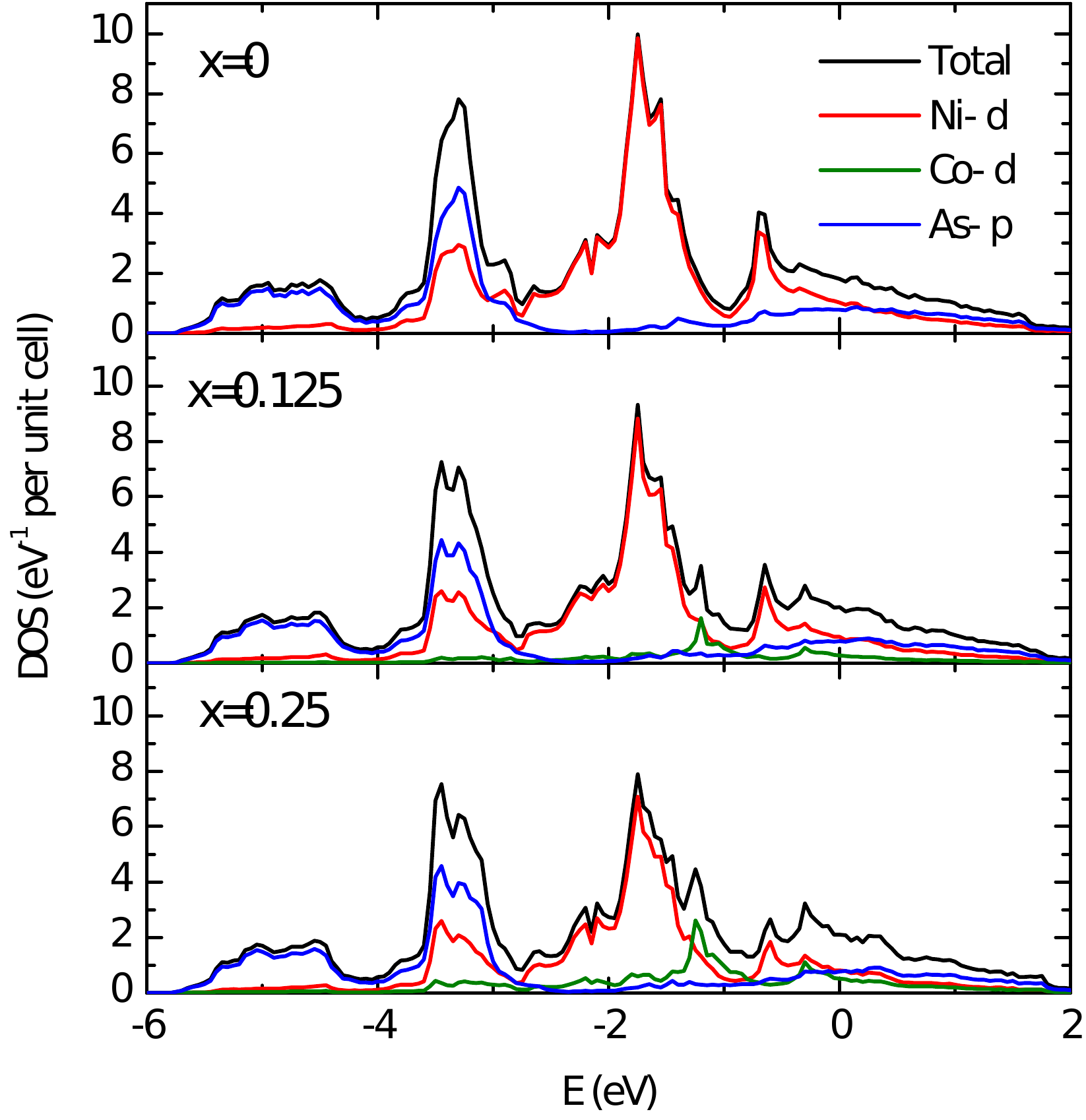}
    \caption{The evolution of the electronic density of states in \BaNiCo. Data displayed for $x$ = 0 (top), 0.125 (center), and 0.25 (bottom).}
    \label{fig:Figure8}
\end{figure}

Previous reports of chemical substitution in \BaNi~feature nearly identical evolutions of superconductivity. However, this trend is broken through Co substitution, which causes a strong enhancement of \Tc~{\it within} the triclinic phase and a smooth evolution through the triclinic-tetragonal T=0 boundary, manifesting in a superconducting dome-shaped phase diagram (Fig. 7). 

The calculated electronic density of states (DOS) exhibits a monotonic enhancement at $E_F$ with increasing Co concentration (Fig. 8), with a Co d-orbital component that smoothly adds to the total DOS. This could be expected to provide an environment more hospitable to superconductivity with increasing Co, but the observed suppression of \Tc~with high Co concentration is inconsistent with this conclusion, ruling out changes in DOS as the predominant factor responsible for enhanced \Tc. We also observe no dramatic changes to Fermi surface topology that account for the rapid suppression of superconductivity in over-substituted \BaNiCo. While the enhanced Wilson ratio observed in the Co-based end-member BaCo$_2$As$_2$ \cite{BaCo} suggests that increasing Co concentration may ultimately invoke ferromagnetic correlations, the concentrations where \Tc~is suppressed in \BaNiCo~series are far from $x$ = 1 such that it is unlikely that the rapid suppression of \Tc~in the tetragonal phase is a result of proximity to ferromagnetism, and warrants further investigation.

Ba(Ni$_{1-x}$Cu$_{x}$)$_2$As$_2$ and BaNi$_2$(As$_{1-x}$P$_{x}$)$_2$ series also exhibit phonon softening in high \Tc~samples, indicated by strong superconducting coupling and a dramatically reduced Debye temperature. While strong superconducting coupling is also observed in the \BaNiCo~series, the Debye temperature remains virtually constant over the range of $x$ studied here. Given the strongly first-order nature of the structural transition, phonon softening within the triclinic phase is unexpected, though would not be unprecedented \cite{BaGe-P}. While the effect of phonon softening on pairing near such a strongly discontinuous structural boundary cannot be ignored, the distinct behavior found in the \BaNiCo~series suggests another mechanism is responsible for the strengthening of superconductivity, which appears to be centered around the triclinic-tetragonal critical point. The recent observations of CDW order in \BaNi~\cite{Abbamonte} are a provocative suggestion that the previously mundane view of both superconductivity and the structural distortion in  \BaNi~should be revisited, and that fluctuation-driven superconductivity is a real possibility \cite{Zigzag}. Further, uncovering a new mechanism for superconducting enhancement opens an interesting avenue to potentially extend superconductivity to even higher critical temperatures in this and related systems.

\section{\label{sec:Acknowledge}Acknowledgments}

Research at the University of Maryland was supported by the AFOSR Grant No. FA9550-14-10332 and the Gordon and Betty Moore Foundation Grant No. GBMF4419. We also acknowledge support from the Center for Nanophysics and Advanced Materials as well as the Maryland Nanocenter and its FabLab.

\bibliography{Ba_NiCo_2As2_refs}

\end{document}